\documentclass[acus]{JAC2001}
\usepackage{graphicx}                                                          
\newcommand{\ud}{\mathrm{d}}                                                   
\begin{document}

\onecolumn 
\thispagestyle{empty}

\begin{center}
\begin{tabular}{p{130mm}}

\begin{center}
{\bf\Large
FAST CALCULATION METHODS IN} \\
\vspace{5mm}

{\bf\Large COLLECTIVE DYNAMICAL MODELS}\\
\vspace{5mm}

{\bf\Large OF BEAM/PLASMA PHYSICS}\\

\vspace{1cm}

{\bf\Large Antonina N. Fedorova, Michael G. Zeitlin}\\

\vspace{1cm}

{\bf\large\it
IPME RAS, St.~Petersburg, 
V.O. Bolshoj pr., 61, 199178, Russia}\\
{\bf\large\it e-mail: zeitlin@math.ipme.ru}\\
{\bf\large\it http://www.ipme.ru/zeitlin.html}\\
{\bf\large\it http://www.ipme.nw.ru/zeitlin.html}
\end{center}

\vspace{1cm}

\abstract{ 
We consider an application of modification of our va\-ri\-a\-ti\-o\-nal\--wa\-ve\-let 
approach to some nonlinear collective
model of  beam/plasma physics: Vlasov/Boltzmann-like reduction 
from general BBGKY hierachy
related to modeling of propagation of intense charged particle
beams in high-intensity accelerators and transport systems. 
We use fast convergent
multiscale variational-wavelet representations for solutions which allow to
consider
polynomial and rational type of nonlinearities. The solutions are
represented via the multiscale decomposition in nonlinear high-localized
eigenmodes (waveletons).
In contrast with different approaches we do not use
perturbation technique or linearization procedures. 
}

\vspace{60mm}

\begin{center}
{\large Presented at the Eighth European Particle Accelerator Conference} \\
{\large EPAC'02} \\
{\large Paris, France,  June 3-7, 2002}
\end{center}
\end{tabular}
\end{center}
\newpage

\title{FAST CALCULATION METHODS IN COLLECTIVE DYNAMICAL
MODELS OF BEAM/PLASMA PHYSICS}
\author{Antonina N. Fedorova, Michael G. Zeitlin \\                            
IPME RAS, St.~Petersburg,                                                      
V.O. Bolshoj pr., 61, 199178, Russia                                           
\thanks{e-mail: zeitlin@math.ipme.ru}\thanks{http://www.ipme.ru/zeitlin.html;
http://www.ipme.nw.ru/zeitlin.html}}
\maketitle
\begin{abstract}
We consider an application of modification of our va\-ri\-a\-ti\-o\-nal\--wa\-ve\-let 
approach to some nonlinear collective
model of  beam/plasma physics: Vlasov/Boltzmann-like reduction 
from general BBGKY hierachy
related to modeling of propagation of intense charged particle
beams in high-intensity accelerators and transport systems. 
We use fast convergent
multiscale variational-wavelet representations for solutions which allow to
consider
polynomial and rational type of nonlinearities. The solutions are
represented via the multiscale decomposition in nonlinear high-localized
eigenmodes (waveletons).
In contrast with different approaches we do not use
perturbation technique or linearization procedures. 
\end{abstract}

\section{INTRODUCTION}

We consider applications of numerical--analytical technique based on 
modification of our variational-wavelet approach to nonlinear collective
models of  beam/plasma physics, e.g. some forms of Vlasov/Boltzmann-like reductions 
from general BBGKY hierarchy (section 2).
These equations are related to the modeling of propagation of intense charged particle
beams in high-intensity accelerators and transport systems [1], [2]. In our
approach we use fast convergent
multiscale variational-wavelet representations, which allows to
consider polynomial and rational type of nonlinearities [3]-[16], [17]. 
The solutions are
represented via the multiscale decomposition in nonlinear high-localized
eigenmodes (generalized Gluckstern modes, in some sense), which corresponds to the full
multiresolution expansion in all underlying hidden time/space or
phase space scales. In contrast with different approaches we don't use
perturbation technique or linearization procedures. 
In section 3 after formulation of key points we consider another variational approach
based on ideas of para-products and nonlinear approximation
in multiresolution approach, which gives the possibility
for computations in each scale separately [18].
We consider representation (4) below, where 
each term corresponds to the contribution from the 
scale $i$ in the full underlying
multiresolution decomposition
as multiscale 
generalization of old (nonlinear) $\delta F$ approach [1].
As a result, fast scalar/parallel
modeling demonstrates appearance of high-localized coherent structures (waveletons)
and pattern formation
in systems with complex collective behaviour.

\section{VLASOV/BOLTZMANN--LIKE REDUCTIONS}

Let M be the phase space of ensemble of N particles ($ {\rm dim}M=6N$)
with coordinates
$x_i=(q_i,p_i), \quad i=1,...,N, \quad
q_i=(q^1_i,q^2_i,q^3_i)\in R^3,\quad
p_i=(p^1_i,p^2_i,p^3_i)\in R^3$
with distribution function
$D_N(x_1,\dots,x_N;t)$
and
\begin{eqnarray}
F_N(x_1,\dots,x_N;t)=\sum_{S_N}D_N(x_1,\dots,x_N;t)
\end{eqnarray}
be the N-particle distribution functions ($S_N$ is permutation group of N elements).
For s=1,2 we have from general BBGKY hierarchy:  
\begin{eqnarray}
&&\frac{\partial F_1(x_1;t)}{\partial t}+\frac{p_1}{m}\frac{\partial}{\partial q_1}
F_1(x_1;t)\\
&&=\frac{1}{\upsilon}\int\ud x_2L_{12} F_2(x_1,x_2;t)\nonumber
\end{eqnarray}
\begin{eqnarray}
&&\frac{\partial F_2(x_1,x_2;t)}{\partial t}+\Big(\frac{p_1}{m}
\frac{\partial}{\partial q_1}+\frac{p_2}{m}\frac{\partial}{\partial q_2}-L_{12}\Big)\\
&& F_2(x_1,x_2;t)
=\frac{1}{\upsilon}\int\ud x_3(L_{13}+L_{23})F_3(x_1,x_2;t)\nonumber
\end{eqnarray}
where partial Liouvillean operators are described in [17].
We are interested in the cases when
$$
F_k(x_1,\dots,x_k;t)=\prod^k_{i=1}F_1(x_i;t)+G_k(x_1,\dots,x_k;t),
$$
where $G_k$ are correlation patterns, really have additional reductions 
as in case of Vlasov-like systems.
Then we have in (2), (3) polynomial type of nonlinearities (more exactly, multilinearities).

\section{MULTISCALE ANALYSIS}

Our goal is the demonstration of advantages of the following representation
\begin{equation}
F=\sum_{i\in Z}\delta^i F,
\end{equation}
for the full exact solution for the systems related to equations 
(2), (3). It is possible to consider (4) as multiscale 
generalization of old (nonlinear) $\delta F$ approach [1].
In (4) each $\delta^i F$ term corresponds to the contribution from the 
scale $i$ in the full underlying
multiresolution decomposition 
\begin{equation}
\dots\subset V_{-1}\subset V_0\subset V_1\subset V_2\dots
\end{equation}
of the proper function space ($L^2$, Hilbert,
Sobolev, etc) to which $F$ is really belong.
It should be noted that (4) doesn't based neither on perturbations nor on 
linearization procedures.
Although usually physicists, who prefered computer modelling as a main 
tool of understanding of physical reality, don't think about underlying 
functional spaces, but many concrete 
features of complicated complex dynamics are really 
related not only to concrete 
form/class of operators/equations but also depend on the proper choice of function 
spaces where operators actully act.
Moreover, we have for arbitrary $N$ in the finite N-mode approximation
\begin{equation}
F^N=\sum^N_{i=1}\delta^i F
\end{equation}
the following more useful decompositions:
\begin{eqnarray}
\{F(t)\}=\bigoplus_{-\infty<j<\infty} W_j\quad {\rm or}\quad
\{F(t)\}=\overline{V_0\displaystyle\bigoplus^\infty_{j=0} W_j},
\end{eqnarray}
in case when $V_0$ is the coarsest scale of resolution and where 
$
V_{j+1}=V_j\bigoplus W_j
$
and bases in scale spaces $W_i(V_j)$ are generated from base functions $\psi(\varphi)$
by action of affine group of translations and dilations 
(the so called ``wavelet microscope'').
The following constructions based on variational approach provide 
the best possible
fast convergence properties
in the sense of combined norm  
\begin{equation}
\|F^{N+1}-F^{N}\|\leq\varepsilon
\end{equation}
introduced in [17].
Our five basic points after functional space choice are:
\begin{enumerate}
\item Ansatz-oriented choice of the (multi\-di\-men\-si\-o\-nal) ba\-ses
related to some po\-ly\-no\-mi\-al tensor algebra. Some example related
to general BBGKY hierarchy is considered in [17].
\item The choice of proper variational principle. A few 
pro\-je\-c\-ti\-on/ \-Ga\-ler\-kin\--li\-ke 
principles for constructing (weak) solutions are considered in [3] - [16].
It should be noted advantages of formulations related to biorthogonal (wavelet) decomposition.
\item
The choice of  bases functions in scale spaces $W_j$ from wavelet zoo. They 
correspond to high-localized (nonlinear) oscillations/excitations, 
coherent (nonlinear) resonances,
etc. Besides fast convergence properties of 
the corresponding variational-wavelet expansions it should be noted 
minimal complexity of all underlying calculations, especially in case of choice of wavelet
packets which minimize Shannon entropy. 
\item Operators  representations providing maximum sparse representations 
for arbitrary (pseudo) differential/ integral operators 
$\ud f/\ud x, \ud^n f/\ud x^n, \int T(x,y)f(y)\ud y)$, etc [17].
\item (Multi)linearization. Besides variation approach we consider now a different method
to deal with (polynomial) nonlinearities.
\end{enumerate}
We modify the scheme of our variational
approach in such a way in which we consider different scales of multiresolution decomposition
(5) separately. For this reason
we need to compute errors of approximations. The main problems come of course
from nonlinear (polynomial) terms. We follow according to the multilinearization
(in case below -- bilinearization) approach 
of Beylkin, Meyer etc from [18].
Let $P_j$ be projection operators on the subspaces $V_j$ (5):
\begin{equation}
(P_j f)(x)=\sum_k <f,\varphi_{j,k}> \varphi_{j,k}(x)
\end{equation}
and $Q_j$ are projection operators on the subspaces $W_j$:
$
Q_j=P_{j-1}-P_j
$.
So, for $u\in L^2(R)$ we have $u_j=P_ju\quad$ and $u_j\in V_j$.
It is obviously that we can represent $u_0^2$ in the following form:
\begin{equation}\label{eq:form1}
u_0^2=2\sum^n_{j=1}(P_ju)(Q_ju)+\sum^n_{j=1}(Q_ju)(Q_ju)+u_n^2
\end{equation}
In this formula there is no interaction between different scales.
We may consider each term of (\ref{eq:form1}) as a bilinear mappings:
\begin{eqnarray}\label{eq:form2}
\displaystyle
M_{VW}^j : V_j\times W_j\to L^2({\bf R})=
V_j{\oplus_{j'\geq j}W_{j'}}
\end{eqnarray}
\begin{eqnarray}\label{eq:form3}
M_{WW}^j : W_j\times W_j\to L^2({\bf R})=V_j\oplus_{j'\geq j}W_{j'}
\end{eqnarray}
For numerical purposes we need formula (\ref{eq:form1}) with a finite number of
scales, but when we consider limits $j\to\infty$ we have
\begin{equation}
u^2=\sum_{j\in {\bf Z}}(2P_ju+Q_ju)(Q_ju),
\end{equation}
which is para-product of Bony, Coifman and Meyer [18].
Now we need to expand (\ref{eq:form1}) into the wavelet bases. To expand
each term in (\ref{eq:form1}) we need to consider
the integrals of the products of the basis functions (7), e.g.
\begin{equation}
M^{j,j'}_{WWW}(k,k',\ell)=\int^\infty_{-\infty}\psi^j_k(x)
\psi^j_{k'}(x)\psi^{j'}_\ell(x)\ud x,
\end{equation}
where $j'>j$ and
\begin{equation}
\psi^j_k(x)=2^{-j/2}\psi(2^{-j}x-k)
\end{equation}
are the basis functions (7).
For compactly supported wavelets 
\begin{equation}
M_{WWW}^{j,j'}(k,k',\ell)\equiv 0\quad \mbox{for}\quad |k-k'|>k_0,
\end{equation}
where $k_0$ depends on the overlap of the supports of the basis functions
and
\begin{equation}\label{eq:form4}
|M_{WWW}^r(k-k',2^rk-\ell)|\leq C\cdot 2^{-r\lambda M}
\end{equation}
Let us define $j_0$ as the distance between scales such that for a given
$\varepsilon$ all the coefficients in (\ref{eq:form4}) with labels
$r=j-j'$, $r>j_0$ have absolute values less than $\varepsilon$. 
\begin{figure}[htb]                                                            
\centering                                                                      
\includegraphics*[width=60mm]{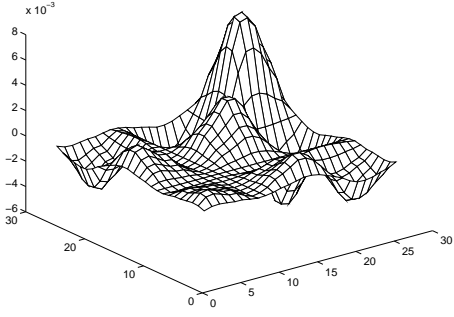}                              
\caption{$N=1$ waveleton contribution to (6).}                                          
\end{figure}                                                                    
\begin{figure}[htb]                                                     
\centering                                                                      
\includegraphics*[width=60mm]{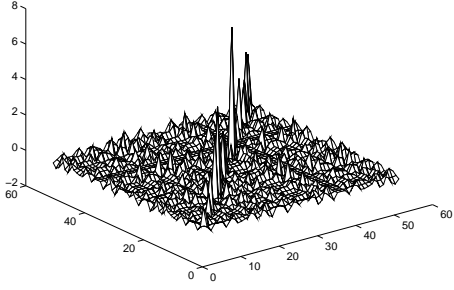}
\caption{Stable pattern.}
\end{figure}
For the purposes of
computing with accuracy $\varepsilon$ we replace the mappings in
(\ref{eq:form2}), (\ref{eq:form3}) by
\begin{equation}\label{eq:z1}
M_{VW}^j : V_j\times W_j\to V_j\oplus_{j\leq j'\leq j_0}W_{j'}
\end{equation}
\begin{equation}\label{eq:z2}
M_{WW}^j : W_j\times W_j\to V_j\oplus_{j\leq j'\leq j_0}W_{j'}\nonumber
\end{equation}
Since
$
V_j\oplus_{j\leq j'\leq j_0}W_{j'}=V_{j_0-1},
\ 
V_j\subset V_{j_0-1},\ W_j\subset V_{j_0-1}
$
we may consider bilinear mappings (\ref{eq:z1}), (\ref{eq:z2}) on
$V_{j_0-1}\times V_{j_0-1}$.
For the evaluation of (\ref{eq:z1}), (\ref{eq:z2}) as mappings
$V_{j_0-1}\times V_{j_0-1}\to V_{j_0-1}$
we need significantly fewer coefficients than for
mappings (\ref{eq:z1}), (\ref{eq:z2}). It is enough to consider only
coefficients
\begin{equation}
M(k,k',\ell)=2^{-j/2}\int^\infty_\infty\varphi(x-k)\varphi(x-k')\varphi(x-\ell)\ud
x,
\end{equation}
where $\varphi(x)$ is scale function. Also we have
\begin{equation}
M(k,k',\ell)=2^{-j/2}M_0(k-\ell,k'-\ell),
\end{equation}
where
\begin{equation}
M_0(p,q)=\int\varphi(x-p)\varphi(x-q)\varphi(x)\ud x
\end{equation}
$M_0(p,q)$
satisfy the standard system of linear equations and after its solution 
we can recover all bilinear quantities (14). Then we apply some variation
approach from [3]-[16], but in each scale separately.
So, after application of points 1-5 above, we arrive to explicit 
numerical-analytical realization of
representations (4) or (6). Fig.1 demonstrates the first contribution to
the full solution (6) while Fig.2 presents (stable) pattern as solution of system
(2)-(3). We evaluate accuracy of calculations according to norm introduced in    
[17].

 \end{document}